\documentstyle[12pt]{article}
\begin{document}
\begin{minipage}[center]{9cm}
\\ \\
{\bf  \Large
Ferron-like states in  YBa$_2$Cu$_3$O$_{6+x}$} \\ \\
P.~Rubin and A.~Sherman\\ \\
{\it  Institute of Physics, University of Tartu, Riia 142,
\newline EE-2400 Tartu,
Estonia}
\end{minipage}
\\  \\   \\  \\

\noindent
With the use of the Hubbard model bound
hole states in $\rm YBa_2Cu_3O_{ 6+ \it x}$ are studied.
For the parameters of this crystal the exchange interaction between
the spin-carrying chain ion $\rm O^-$ and $\rm Cu-O$ plane sites is
shown to ensure the formation of a large ferromagnetically ordered
clusters
around holes
in the plane.    \\ \\


\noindent   {\Large\bf  1.~Introduction}               \\  \\

\noindent Over
the past decade the magnetic structure of cuprate perovskites has been
studying very intensively in the hope that it might provide insight in
the physical origin of high-temperature superconductivity. It is
presently well established that $\rm CuO_2$ planes of undoped and
lightly doped cuprates are antiferromagnetically ordered. Some
experiments in such crystals, for example magnetization measurements
in  $\rm YBa_2Cu_3O_{ 6+ \it x}$  \cite{ratner}, suggest the existence
of local ferromagnetically ordered regions around holes on the
antiferromagnetic background of $\rm CuO_2$ planes. In application to
another type of crystals such excitations were first considered in
Ref.~2
and were called ferrons. Ferron models of high-$T_c$ materials were
discussed in Refs.~3,4. The explanation of the ferron formation,
suggested
in Ref.~1, is based on the comparison of antiferromagnetic and
ferromagnetic interactions created by a hole localized on an oxygen
orbital. This idea goes back to work \cite{aharony}
where it was used for the explanation  of the  observed destruction
of the long-range antiferromagnetic order at small hole concentrations.
The mechanism suggested in the latter work is essentially based on the
supposition that nonhybridized oxygen and copper orbitals are
appropriate starting states for the consideration.  However as shown
in Ref.~6, low-lying  states are strongly hybridized. As a consequence,
the mechanism of Ref.~5 is not appropriate for explaining the
destruction  of the long-range ordering. For similar reasons this
mechanism  may not be responsible for the ferron formation.

Notice that the mentioned strong hybridization of oxygen and copper
orbitals and the crystal structure of $\rm CuO_2$    planes open the
way to reduce the Emery Hamiltonian for these planes to the more simple
$t$-$J$ or Hubbard Hamiltonians with only one effective orbital per
plane
sell \cite{rice}. In the $t$-$J$ model the ferron formation was
considered both by the exact diagonalization of small clusters
\cite{bonca,moreo,hasegava} and in an infinite plane with the use of
the spin-wave approximation \cite{ShermanI,ferron}.
Both approaches showed that in the periodic lattice ferrons are
energetically unfavorable for parameters of cuprate perovskites.
The next step is to check whether point defect can stabilize ferrons.
In our recent paper \cite{rushsch} we have considered local states
generated by the interstitial $\rm O^{2-}$
ion in $\rm La_2CuO_{4+\delta}$
and found that this type of defects is unable to stabilize ferrons.
In  $\rm YBa_2Cu_3O_{ 6+ \it x}$,  where ferrons  were experimentally
observed, for small $x$
oxygen ions start to fill chain sites and
can be considered as interstitial
defects above the $\rm CuO_2$ plane (see Fig.~1). Their valency is
equal to -2 or -1 \cite{uimin} and in the latter case the defect has
the spin 1/2. It can be expected that, in contrast to the spinless
$\rm O^{2-}$ ion, such type of defects  can stabilize states with a
large
spin, i.e. ferrons, due to an arising exchange interaction.
\\ \\

\noindent {\Large\bf  2.~Basic Formulae}  \\  \\

\noindent
To check this idea we use the one-band Hubbard model for the
description
of a $\rm CuO_2$ plane. The model Hamiltonian reads

\begin{equation}
H_{\rm 0} = U \sum_{\bf l} n_{{\bf l}\uparrow} n_{{\bf l}\downarrow}
- t \sum_{\bf l a \sigma} c^\dagger_{\bf l \sigma}
c_{\bf l+a \sigma},
\end{equation}
\noindent where the operator $c^\dagger_{\bf l \sigma}$ creates an
 electron in the effective orbital on the plane
site ${\bf l}$ (the orbitals are centered at copper sites)
with the spin  $\sigma$,
${\bf a} = (0,\pm a),(\pm a,0)$, where $a$ is the in-plane copper
distance,
$n_{{\bf l}\sigma}=c^\dagger_{\bf l \sigma}
c_{\bf l \sigma}$,
$U$ and $t$ are the Coulomb repulsion and the hopping matrix element,
respectively.
A chain oxygen $\rm O^- $ is situated at ${\rm O1}$ position and we add
 the
following terms to the Hamiltonian
\begin{equation}
H_{\rm 1} = \Delta \sum_{\sigma} a^\dagger_\sigma a_\sigma  -
 \sum_{\bf m \sigma} t_{\bf m} \left(a^\dagger_\sigma c_{\bf m \sigma}
+ c^\dagger_{\bf m \sigma} a_{\sigma } \right)
+ U_p a^\dagger_{\uparrow} a_{\uparrow}
a^\dagger_\downarrow a_\downarrow ,
\end{equation}
\noindent where $a^\dagger_\sigma$ is the $p$-electron creation
operator in the ${\rm O^-}$ chain site, $\Delta$ is the difference in
chain  and  plane site energies,
$U_p$ is the Coulomb repulsion on the chain oxygen,
and $t_{\bf m}$  is the hopping
matrix element between chain and plane sites.
Summation over $\bf m$ proceeds over the plane sites in the vicinity
of the ``defect'' chain oxygen site (these plane sites are shown in
Fig.~1) and we suppose that $t_{\bf m} \approx t$ for these sites.

For  $\rm YBa_2Cu_3O_{ 6+ \it x}$ the parameters in Eqs.~(1) and (2) can
be estimated as     $t_{\bf m} \approx t \approx 1 \, \rm eV$,
$\Delta \approx -2 \rm \, eV$,  $U \approx 10 \,
{\rm eV}, U_p \approx 4.5 \, \rm eV$  \cite{ShermanI,mahan}.
For such parameters the correlation
 terms in Eqs.~(1),(2)
and the first term in Eq.~(2)
can be used as the
unperturbed
Hamiltonian  and the second, hopping terms, as the perturbation in the
perturbation theory expansion. The resulting effective Hamiltonian reads
$$
H_{\rm eff} = P \left(
- t \sum_{\bf l a \sigma}  c^\dagger_{\bf l \sigma}
c_{\bf l+a \sigma} + \frac{J}{2}
\sum_{\bf l a} {\bf S}_{\bf l+a}
{\bf S}_{\bf l}
+
\sum_{\bf m} J_{\bf m} {\bf S}_{\bf m}
{\bf S}_{\bf o}
\right) P,
$$
\begin{equation}
\quad \quad \quad \quad \quad \quad \quad \quad
J = \frac{4 t^2}{U}, \; \quad
J_{\bf m} = \frac{2 t_{\bf m}^2 (U+U_p)}{(\Delta + U_p) (U - \Delta)},
\end{equation}
where
$$
S_{\bf l}^z = \sum_{\sigma} \sigma c_{\bf l \sigma}^\dagger
c_{\bf l \sigma}/2,    \quad
S^\sigma_{\bf l }= S_{\bf l}^x + i \sigma S_{\bf l}^y =
c_{\bf l \sigma}^\dagger c_{\bf l -\sigma},
$$
are the components of the spin operators $\bf S_l$ on the plane sites.
The analogous Hamiltonian was derived in  \cite{Fulde}.
The components of the
spin operator $\bf S_o $ on the chain site
are defined analogously through
the operators $a_\sigma$.
$P$ is the projection operator on the subspace where the chain
level is singly occupied and the plane levels are singly occupied or
empty.
The first two terms in Eq.~(3) represent
the one-band $t$-$J$ Hamiltonian on a square lattice.
The summation over $\bf l$  proceeds over all plane sites,
while the
summation over $\bf m$, over
the
``defect'' plane sites.
In Eq.~(3), the ``defect'' term  can be
represented as the sum of  terms with $S^z$
and $S^{\pm}$ spin components. The term
with $S^z$  components plays the main role in the ferron stabilization,
whereas the terms
with $S^{\pm}$ components lead only to additional lowering
of energy  levels. For this reason for simplicity we take into account
only components with $S^z$ in the ``defect'' term in Eq.~(3).
Let us suppose
that
the $\rm O1$ site is
occupied by an electron with $S^z=1/2$.
In this case
the effective Hamiltonian can be rewritten in the form
\begin{equation}
H_{\rm eff} =
- t \sum_{\bf l a \sigma} c^\dagger_{\bf l \sigma}
c_{\bf l+a \sigma} + \frac{J}{2}
\sum_{\bf l,a} {\bf S}_{\bf l+a}
{\bf S}_{\bf l}
+ \frac{1}{2}
\sum_{\bf m} J_{\bf m}  S^z_{\bf m}.
\end{equation}
\noindent Since the sign of $J_{\bf m}$ is the same for all sites in
the ``defect'' region, the last term can stabilize ferrons.

We use the
spin-wave approximation \cite{ShermanI,schmitt}
for the description of spin excitations in the plane.
The approximation
reduces to neglecting terms of the third and higher
orders in the spin-wave operators $b_{\bf m}$
introduced into Eq.~(4) by the Holstein-Primakoff formulas
$$  
S_{\bf l}^{+1}  = \Phi_{\bf l} b_{\bf l} P_{\bf l}^{+1} +
 b_{\bf l}^\dagger \Phi_{\bf l} P_{\bf l}^{-1},\;   \;
S_{\bf l}^{-1} =
 (S_{\bf l}^{+1})^\dagger , \; \;
  S_{\bf l}^{z} = e^{i{\bf \Pi l}} \left(
{\sum_{\sigma} n_{{\bf l} \sigma}\over2}-b_{\bf l}^\dagger
b_{\bf l} \right),
$$ 
\noindent
where
$\Phi_{\bf l}  =  \left(1-b_{\bf l}^\dagger b_{\bf l} \right)^{1/2} ,$
${\bf \Pi}=(\pi /a,\pi /a), \;         {\rm and}  \;
P_{\bf l}^\sigma = [1+\sigma \exp (i {\bf \Pi l})]/2$.

Let us introduce  the  hole  creation  operator
$h_{\bf l}^{\dagger} =$ $ \sum_{\sigma}
P_{\bf l}^{\sigma}
c_{\bf l \sigma}
$
for the N\'eel state
 $| {\cal N} \rangle.$
This classical N\'eel state can be determined by the condition
$b_{\bf l} | {\cal N} \rangle = 0$
(for the second N\'eel state
analogous formulas are
obtained by substituting
$P^\sigma_{\bf l}$ with $1 - P^\sigma_{\bf l}$).
After the  unitary transformation which
diagonalizes  the Heisenberg part of the Hamiltonian  \cite{ShermanI}
(the second term in Eq.~(4))
the effective Hamiltonian acquires the form
\begin{eqnarray}
H_{\rm eff}&=& \sum_{\bf l l^{\prime} a}
t
\left[
h_{\bf l+a}
h_{\bf l}^{\dagger}
b_{\bf l - l^{\prime}}^{\dagger}
(u_{\bf l^{\prime} + a} + v_{\bf l^{\prime}})
 + {\rm H.c.}
 \right]
+ \frac{J}{2} \sum_{\bf l l^{\prime}}\omega_{\bf l^{\prime}}
 b_{\bf l}^\dagger b_{\bf l + l^{\prime}}   \nonumber   \\
&-& \sum_{\bf m} \left( J_{\bf 1 m}  \, \tilde{n}_{\bf m}
+  J_{\bf 2 m}  \,         b^\dagger_{\bf m} b_{\bf m}
\right) \exp (i {\bf \Pi m}),
\end{eqnarray}
\noindent where
$u_{\bf l},\: v_{\bf l}$ and $\omega_{\bf l}$
are the Fourier transforms of $\cosh (\alpha_{\bf k}),\:
-\sinh (\alpha_{\bf k})$, and
$ \omega_{\bf k } = 4 \sqrt{1 - \gamma^2_{\bf k}}$,
respectively,
$
\alpha_{\bf k}=( 1 / 8 ) \, \ln \left[ (1+\gamma_{\bf k} ) /
(1-\gamma_{\bf k} ) \right] ,\; $
$\gamma_{\bf k} = $ \newline $ [ \cos{(k_xa)}$ $ + \cos{(k_ya)} ] / 2$,
$\tilde{n}_{\bf m} = h^{\dagger}_{\bf m}   h_{\bf m}$.
Due to the fast decrease of $u_{\bf l^{\prime} +a}
+ v_{\bf l^{\prime}}$
and $\omega_{\bf l^{\prime}}$ with the
growth of $\vert {\bf l^{\prime}}
\vert$ only the components with $\bf l^{\prime}
$ $= ( \pm a, 0), (0, \pm a) $ for the sum in
the first, kinetic energy term and
the components with ${\bf l^{\prime}} =
(0,0),(\pm a, \pm a),(\pm 2 a,0),(0,\pm 2 a) $
for $\omega_{\bf l^{\prime}}$
in the second, magnetic energy term are retained
in Eq.(5)
in the subsequent calculations.
In this equation,
$J_{\bf 1 m} = J_{\bf m} / 4,$ $J_{\bf 2 m} = 2 (
u_{\bf 0}^2 +
v_{\bf 0}^2)
J_{\bf 1 m}   $ where $
u_{\bf 0}^2 +
v_{\bf 0}^2
\approx 1.16$.

To investigate the lowest eigenvalues and eigenfunctions of
Hamiltonian (5) we  use the modified Lanczos algorithm
\cite{ShermanI,Dagotto}. In one step of this algorithm we determine a
final
state $\vert {\bf f} \rangle $ from an initial state
$\vert {\bf i} \rangle $
according to the procedure
\begin{equation}
\quad \quad
\langle {\bf i} \vert {\bf i} \rangle = 1, \; \; {\cal E}_i =
\langle {\bf i} \vert {\cal H} \vert {\bf i} \rangle,
\; \; {\cal V} \vert
{\bf f} \rangle = ({\cal H - E}_i) \vert {\bf i} \rangle .
\end{equation}

\noindent Here ${\cal V}$ is the normalization constant which is
determined from
the condition $ \langle {\bf f}
\vert {\bf f} \rangle =1. $ After one Lanczos step
the energy is minimized in the subspace of the states
$ \vert {\bf i} \rangle $ and $ \vert {\bf f} \rangle $
(it follows from (6)  that                $
\langle {\bf i} \vert {\bf f} \rangle = 0 )$.
The obtained
function
$ \vert {\bf i^\prime} \rangle = c_1 \vert {\bf i} \rangle +
c_2 \vert {\bf f} \rangle $ which minimizes the energy
is used as the initial function in the next Lanczos step.
This procedure is continued until the ne\-ces\-sa\-ry accuracy of the
 eigenvalue is reached. If the initial function $\vert \bf i \rangle$
is characterized by some spin projection $S^z$, $\vert {\bf f} \rangle$
also corresponds to the same $S^z$. Therefore, it is possible to
obtain the lowest eigenvalues of  Hamiltonian (8) with
a given $S^z$ by  starting from a state with the respective $S^z$.
Notice that states with  $S^z > 1/2$ contain flipped spins near the
hole and correspond to ferrons.

The number of spin configurations which is necessary for describing
the states $ \vert {\bf i} \rangle$ and
$\vert {\bf f} \rangle$  grows from step to step. To avoid the
overflow of the computer storage and to reduce  the
computation
time,
in each Lanczos step
the state $\vert {\bf i^\prime}
\rangle$ is restricted  to  the {\it j} configurations with the largest
amplitudes \cite{ShermanI}. After normalization this restricted state
is used as the
initial state for
the next
recursion step
of the Lanczos procedure (6). We used {\it j} = 200.
The $\vert {\bf f} \rangle$ state obtained from such
$\vert {\bf i^\prime} \rangle$ state and used for the calculation of the
eigenvalue contains of the order of $10^4$ spin configurations.
This is comparable to the number of configurations used in exact
diagonalization studies of small lattices. To check the accuracy of
the obtained results in some cases we increased {\it j} up to 250 which
led only to small quantitative changes (of the order of
$10^{-4}t $) in
the eigenvalue.    \\  \\

\newpage

\noindent {\Large\bf 3.~Results and Discussion}  \\  \\

\noindent
Using the above-described algorithm we have calculated the lowest
one-hole
states characterized by $S^z \le 13/2.$  For parameters of
cuprate perovskites the ratio $J/t$ can be estimated to lie in the
range
0.2 - 0.5  \cite{ShermanI}.
According to Eq.~(3) the value of $J_{\bf m} /t$
for nearest neighbours  of O1  oxygen site 
can be estimated to be equal 1. For other sites in the  ``defect''
region the values of $J_{\bf m}/t = 0.7$ is accepted.
The results of
our calculations for the lowest one-hole eigenvalues
are shown in Fig.~2.
As seen,  the interaction between the $\rm O^-$ ion spin with
plane spins leads to the ferron formation: the lowest state has
$S^z = 7/2$.
The optimal ferron size can be expected to grow for two and larger
number of adjacent $\rm O^-$ ions in a chain.

In summary, with the use of the spin-wave approximation and the
Lanczos algorithm the localized hole states induced by  an
$\rm O^-$ chain ion in $\rm YBa_2Cu_3O_{\rm 6+ \it x}$
were investigated.
The lowest hole state is characterized by the $z$ component of the
total spin $S^z = 7/2$ which corresponds to a comparatively large
ferromagnetic region with 3 overturned spins around a hole. This
means that the spin-carrying $\rm O^-$ ion can stabilize ferrons.

{\it Acknowledgements}  This work was supported by the Estonian
Science Foundation under Grant ETF-2688.

\newpage

Figure captions. \\

Fig.~1.  The ``defect'' region in the vicinity of the chain
$\rm O^-$  ion (O1).

Fig.~2.  The dependence of energies of the localized hole states
induced by the $\rm O^-$  ion on $S^z$.


\begin{thebibliography}{26}
 \bibitem{ratner}
S.~L.~Gnatchenko, A.~M.~Ratner, M.~Baran, R.~Szymczak,
 and H.~Szymczak,   Phys. Rev. B {\bf 55}, 3876
(1997).
 \bibitem{gennes} P.~G.~de~Gennes,
J. Phys. Radium {\bf 23}, 630 (1962).
 \bibitem{dkhv}   V.~L.~Pokrovskii, G.~V.~Uimin, and
D.~V.~Khveshchenko,
JETP Lett. {\bf 46 Suppl.}, S113 (1987).
 \bibitem{hizhnyakov} V.~Hizhnyakov and E.~Sigmund,  Physica
 C {\bf 156}, 655
 (1988).
\bibitem{aharony} R.~J.~Birgeneau, M.~A.~Kastner, and A.~Aharony,
 Z.~Phys. B {\bf 71}, 57 (1988).
 \bibitem{rice} F.~C.~Zhang and T.~M.~Rice, Phys. Rev.
B {\bf 37}, 3759 (1988).
\bibitem{bonca} J.~Bon\v{c}a, P.~Prelov\v{s}ek, and I.~Sega,
Phys. Rev. B {\bf 39}, 7074 (1989).
\bibitem{moreo} E.~Dagotto, A.~Moreo, and T.~Barnes,
Phys. Rev. B {\bf 40}, 6721 (1989).
\bibitem{hasegava} Y.~Hasegawa and D.~Poilblank,
Phys. Rev. B {\bf 40}, 9035 (1989).
\bibitem{ShermanI}
A.~V.~Sherman,  Solid State Commun. {\bf 76}, 321 (1990),
Phy\-si\-ca C {\bf 171} 395 (1990).
 \bibitem{ferron} J.~Sabczynski, M.~Schreiber, and A.~Sherman,
Phys. Rev.
 B {\bf 48}, 543 (1993).
\bibitem{rushsch} P.~Rubin, A.~Sherman, and M.~Schreiber,
Phys. Rev. B  {\bf 57}, 10299    (1998).
\bibitem{uimin} D.~R.~Grempel, H.~Haugerud, G.~Uimin,
  and W.~Selke, Czechoslovak Journal of Physics
 {\bf 46 Suppl. S2}, 967
 (1996).
\bibitem{mahan} A.~K.~McMahan,  J.~F.~Annett, and
R.~M.~Martin, Phys. Rev. B {\bf 42}, 6268 (1990).
\bibitem{Fulde} T.~Schork and P.~Fulde, Phys. Rev. B {\bf 50},
1345 (1994).
\bibitem{schmitt} S.~Schmitt-Rink, C.~M.~Varma,
and A.~E.~Ruckenstein,
Phys. Rev. Lett. {\bf 60}, 2793 (1988).
\bibitem{Dagotto} E.~Dagotto and A.~Moreo, Phys. Rev.
D {\bf 31}, 865 (1985).
\end{thebibliography}
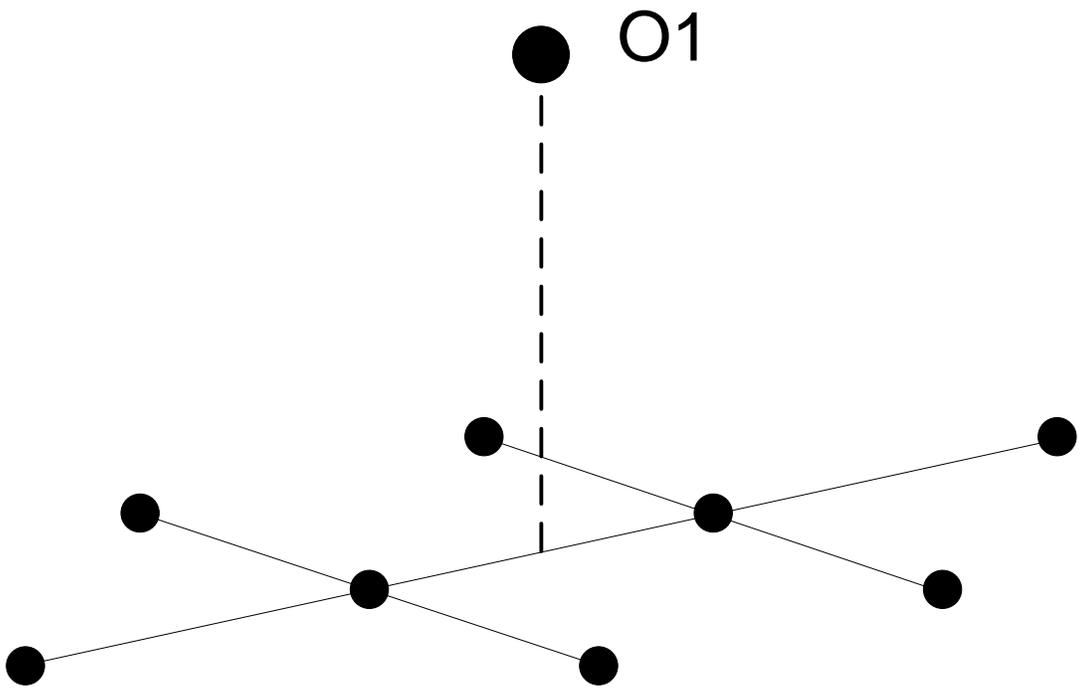
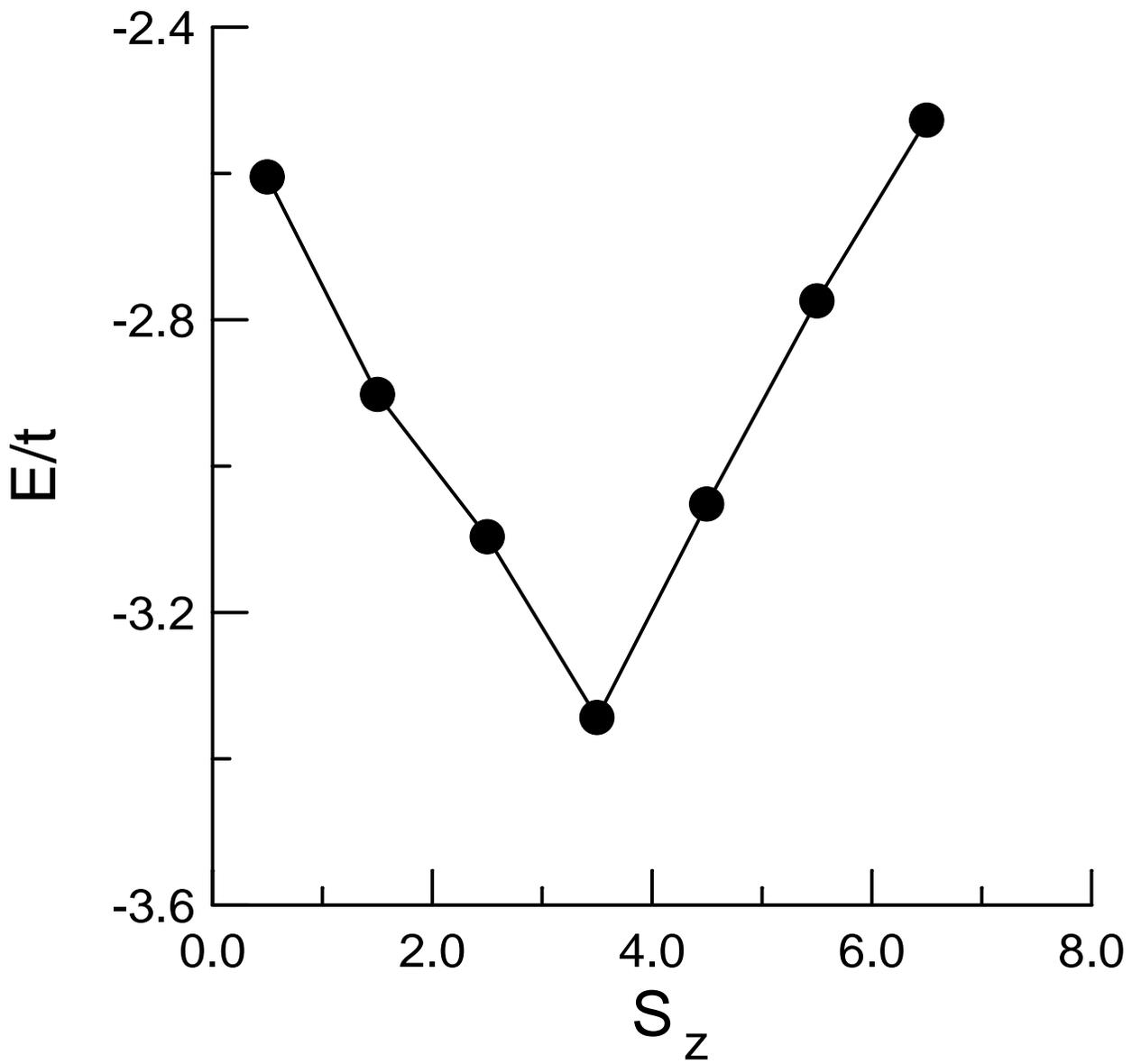
\end{document}